# Sizeable suppression of thermal Hall effect upon isotopic substitution in strontium titanate

Sangwoo Sim[1,2,3]*, Heejun Yang[1,2,3]*, Ha-Leem Kim[2,3], Matthew J Coak[2,3], Mitsuru Itoh[4,5], Yukio Noda[6], and Je-Geun Park[1,2,3]&

1. Center for Quantum Materials, Seoul National University, Seoul 08826, Korea
2. Center for Correlated Electron Systems, Institute for Basic Science, Seoul 08826, Korea
3. Department of Physics & Astronomy, Seoul National University, Seoul 08826, Korea
4. Laboratory for Materials and Structures, Tokyo Institute of Technology, 4259 Nagatsuta, Midori-ku, Yokohama 226-8503, Japan
5. Research Institute for Advanced Electronics and Photonics (RIAEP), National Institute of Advanced Industrial Science and Technology Central-2, 1-1-1 Umezono, Tsukuba, Ibaraki 305-8568, Japan
6. Institute of Multidisciplinary Research for Advanced Materials, Tohoku University, Sendai 980-8577, Japan

* Equal contribution
& Corresponding authors: jgpark10@snu.ac.kr

## Abstract

We report measurements of the thermal Hall effect in single crystals of both pristine and isotopically substituted strontium titanate. We discovered a two orders of magnitude difference in the thermal Hall conductivity between $SrTi^{16}O_3$ and $^{18}O$-enriched $SrTi^{18}O_3$ samples. In most temperature ranges, the magnitude of thermal Hall conductivity ($\kappa_{xy}$) in $SrTi^{18}O_3$ is proportional to the magnitude of the longitudinal thermal conductivity ($\kappa_{xx}$), which suggests a phonon-mediated thermal Hall effect. However, they deviate in the temperature of their maxima, and the thermal Hall angle ratio ($|\kappa_{xy}/\kappa_{xx}|$) shows anomalously decreasing behavior below the ferroelectric Curie temperature $T_c$ ~25 K. This observation suggests a new underlying mechanism, as the conventional scenario cannot explain such differences within the slight change in phonon spectrum. Notably, the difference in magnitude of thermal Hall conductivity and rapidly decreasing thermal Hall angle ratio in $SrTi^{18}O_3$ is correlated with the strength of quantum critical fluctuations in this displacive ferroelectric. This relation points to a link between the quantum critical physics of strontium titanate and its thermal Hall effect, a possible clue to explain this example of an exotic phenomenon in non-magnetic insulating systems.

Heat can be carried by any quasiparticles based on the principle of kinetic theory, which has been long formulated into the Boltzmann transport theory [1]. Unlike the electrical current of electrons, these different carriers of heat, therefore, can probe diverse facets of the underlying rich physics. When charge neutral quasiparticles like magnons, spinons, or phonons become the medium of heat transport, it is, a priori, more challenging to develop a transverse heat gradient under a perpendicular magnetic field [2]. The main reason is that there is simply no obvious candidate for generating a Lorentz-type force on these charge neutral particles.

However, the thermal Hall effect (THE) has now been experimentally observed in several materials, whose ground states cover a wide variety of materials. The examples include paramagnetic [3], ferromagnetic [4], quantum spin ice [5], kagome magnet [6], multiferroic [7], quantum spin liquid & phonon glass system [8], Kitaev spin liquid [9,10], pseudogap phase of cuprates [11], and quantum paraelectric [12] phases. Depending on the ground states, different scenarios have so far been invoked to explain the measured THE. For example, both the observation of THE in $RuCl_3$ and, more importantly, its quantization, have been attributed to the elusive Majorana fermion [9,10]. On the other hand, THE in magnetic systems like $Lu_2V_2O_7$ [4], $Tb_2Ti_2O_7$ [5], Cu(1,3-bdc): bdc stands for benzendicarboxylate [6], $(FeZn)Mo_3O_8$ [7], and $Ba_3CuSb_2O_9$ [8] tend to be interpreted as either direct effects of magnons or indirect consequence via spin-lattice coupling [13-17]. It is also notable that large values of THE were recently observed in high-temperature superconductors and suggested as evidence of chiral phonons [11,18].

The phonon is the most obvious candidate for carrying heat current and most likely the only one for the case of a nonmagnetic insulator. As such, a phonon-based scenario was considered for the THE when it was first observed in paramagnetic $Tb_3Ga_5O_{12}$ [3]. The observed phonon Hall effect in $Tb_3Ga_5O_{12}$ can be formulated in terms of spin-orbit coupling [31] or resonant scattering between phonons and the crystal field level of the $Tb^{3+}$ ion [32]. However, THE in non-magnetic insulating system seems to have no obvious mechanism of how the system couples to external magnetic field. Recent theory papers predicted thermal Hall coefficient ($\kappa_{xy}$) values for non-magnetic band insulator Si crystals [33] and ionic insulator NaCl [29] due to spin-orbit coupling, but these are yet to be verified due to the values predicted being too small to measure.

The latest report of THE in $SrTiO_3$ is unique as none of the three constituent elements are magnetic, and no magnetism has been demonstrated in the material, so there any connection to magnons can be ruled out [12]. Also, the observed thermal Hall conductivity was large compared to previously measured systems, which is especially notable. The authors of Ref. [12] concluded that THE they observed is due to or enhanced by the presence of the antiferrodistortive (AFD) structural domains in $SrTiO_3$.

$SrTiO_3$ is an insulating cubic perovskite with an AFD cubic-tetragonal transition at around 105 K. There has been a long history of broad and detailed literature reporting on its unique properties and a slew of useful applications. $SrTi^{16}O_3$ is an incipient ferroelectric – it has an anomalously high dielectric constant due to its proximity to a long-range ordered ferroelectric state on its phase diagram. Replacing the oxygen with the heavier $^{18}O$ isotope alters the mass of the phonon oscillators and brings their energy to zero, leading to bulk ferroelectric order. SrTi$^{16}$O$_3$ has a small frozen phonon energy gap and has been shown to exhibit ferroelectric quantum critical fluctuations [19, 20]. In this system, all the physics is primarily governed by a single transverse optical phonon mode, removing many complexities and complications otherwise seen in other materials. Additionally, SrTiO$_3$, when doped with a minuscule concentration of charge carriers, exhibits exotic unconventional superconductivity. It is still far from understood and under intense scrutiny worldwide but increasingly linked to this phonon mode and quantum criticality [21, 22].

In this Letter, we report a difference by a factor of 100 times in the thermal Hall conductivity of two types of SrTiO$_3$ samples: one prepared in $^{16}$O and another isotope enriched by preparation in $^{18}$O. With both having the same AFD structural domains forming around 105 and 110 K, respectively [23], it is remarkable that there exists yet this difference in the magnitude of their THE signals. Moreover, a sudden decrease in the thermal Hall angle ratio is discovered in SrTi$^{18}$O$_3$ samples below a certain temperature.

We used SrTiO$_3$ samples with two different oxygen masses. One is a commercial $^{16}$O-based sample from CrysTech, the same as used in Ref. [12]. We cut the sample into a rectangular shape with a dimension of 4.36×2.72×0.5 mm$^3$. Another is $^{18}$O-enriched SrTiO$_3$, and isotope exchange was carried out through the following recipe. A single crystalline SrTi$^{16}$O$_3$ plate (Furuuchi Chemical Co.) with a size of 15×15×0.3 mm$^3$, whose faces were <100> cubic, was used for the exchange. The single-crystalline plate was heated at 1273 K in $^{18}$O$_2$ gas (Isotech Co.) of 99% purity in a quartz glass vessel with an internal volume of 700 cc. The isotope exchange process was repeated several times by exchanging the $^{18}$O$_2$ gas with fresh gas until confirming a weight increase of 3.23% compared to the pristine weight. The total heating time needed to get a constant weight and the 3.23% increase in weight was in excess of six months. The measured capacitance shows a ferroelectric transition near 25 K (see Fig. 1a), indicative of an enrichment level of greater than 98%. We point out that, although unavoidably a small concentration of oxygen sites remain occupied by $^{16}$O in the enriched sample, a similar disorder is in fact present in SrTi$^{16}$O$_3$ samples – they will have $^{18}$O inclusions at a level around its natural abundance of 0.2%. Oxygen site disorder has not been shown, in the detailed literature, to adversely affect ferroelectric order or quantum critical fluctuations, both intrinsically long-ranged and sensitive to disorder and perturbations.

We used a purpose-built THE measurement setup developed in-house for the reported measurements, whose technical details can be found in Ref. [25]. There are several unique features of our instrument. One is that we use three high-quality SrTiO$_3$ crystals from Crystal GmbH as thermometers. These SrTiO$_3$ thermometers provide us with a quick turnaround time by eliminating field-calibration times. We aligned all our single crystals of SrTi$^{18}$O$_3$ (STO18) and SrTi$^{16}$O$_3$ (STO16) to have the <100> axis parallel to the magnetic field, as in Ref. [12]. Custom-made thermal Hall probe with sample puck attached was inserted into the commercial PPMS system, with a temperature range between 15~80 K and magnetic field up to 9 T (see Fig. 1b). We measured both longitudinal ($\kappa_{xx}$), and transverse ($\kappa_{xy}$) thermal conductivity simultaneously by using three thermometers, with thermal conductivities were extracted by symmetrizing and anti-symmetrizing the signals, as $\Delta T_x$ ($\Delta T_y$) is an even (odd) function of magnetic field. Such symmetrizing and anti-symmetrizing process removes from the signals any undesirable effects due to the misalignment of samples. For each measurement temperature, it took typically about 10 hours of saturation time for $\Delta T_y$ signals to get stabilized and come within the window of our experiment resolution. We then waited a further 20~30 minutes after changing magnetic fields before the temperature response saturates. For the measurement of longitudinal thermal conductivity ($\kappa_{xx}$) only, the same system was used with a temperature range between 10~150 K without magnetic field.

As the thermal Hall signals of STO18 samples are tiny compared to STO16, we experimented with the verification of our result. It is essential to show that the thermal Hall signal is not the result of other artefacts like thermal noise and systematic error. We verified an anti-symmetrized $\Delta T_y$ at different thermal gradients and magnetic field values (see Fig. 2). A heater was set to apply a thermal gradient between 3-8% of the average sample temperature. The linearity of $\Delta T_y$ and heater power shows that our signal is distinct from the reference line with zero thermal gradient (noise level) and that the thermal gradient is not excessive – large enough to break the linear nature within its small perturbations. A distinct change with fields in $\Delta T_y$ ensures that the signals are induced by the applied magnetic field. Additionally, the negative sign indicates negative thermal Hall conductivity, which is consistent with what was observed in Ref. [12]. As a reference, we also measured longitudinal thermal conductivity without magnetic field by using a thermal gradient of less than 2% of the average sample temperature.

Our measurements of the longitudinal thermal conductivity ($\kappa_{xx}$) on the commercial STO16 sample are in good agreement with those reported in Ref. [28], providing good verification of our measurement setup (see Fig. 3a). For direct comparison, we then measured two $^{18}$O-enriched STO18 samples to find that their longitudinal thermal conductivity displays a similar temperature dependence to that of the STO16 sample, and so to that in Ref. [28]. The salient point is that while data points above 50 K are more or less the same for all the samples, $\kappa_{xx}$ is suppressed for our two STO18 samples when it is cooled below 50 K and forms a peak.

There are two things we should note here. First, our separate dielectric constant measurement shown in Fig. 1a exhibits a considerable value with a peak at 25 K and a dielectric constant of almost 14,000. The peak in the dielectric constant is indicative of a ferroelectric transition at a Curie Temperature $T_c$, as expected due to the isotope effect. With higher oxygen mass, the zone-center soft transverse optical phonon mode, frozen in STO16, gets condensed, leading to a ferroelectric transition [26]. Another point is that our $^{18}$O specimen also exhibits a plateau in $\kappa_{xx}$ near 110 K, where the AFD structural transition occurs, driven by the condensation of a zone-boundary soft phonon mode. Therefore, our STO18 samples exhibit both the large dielectric constant and the AFD transition, just as in STO16.

Despite the similarity in the longitudinal thermal conductivity $\kappa_{xx}$ between both $^{16}$O and $^{18}$O SrTiO$_3$ samples, there exists a difference by two orders of magnitude in the transverse Hall conductivity $\kappa_{xy}$, as shown in Fig. 3b. Firstly, our STO16 sample shows similar quantitative values of the transverse Hall conductivity to those reported in Ref. [12]. Additionally, the transverse Hall conductivity measured on two STO18 samples shows a somewhat similar temperature dependence as in the STO16 data, showing a peak at a certain temperature. However, despite the qualitatively similar behavior, the $\kappa_{xy}/T$ values at the peak get suppressed by as much as two orders of magnitude as compared with those of STO16. We note additionally that the transverse Hall conductivity shows a linear field dependence for both samples across the range of measured temperatures, as shown in Fig. 3c.

To understand the origin of this drastic isotope effect, we have plotted $\kappa_{xx}/T$ and $\kappa_{xy}/T$ of STO18 together to find that both data show a peak at similar, but seemingly distinct, temperatures of 20 and 25 K, respectively, in contrast to the behavior seen in STO16 presented in Ref. [12] (see Fig. 4a). This observation implies that the similar temperature dependence seen in $\kappa_{xx}/T$ and $\kappa_{xy}/T$ is generic despite the large, two orders of magnitude, difference in $\kappa_{xy}$ between the $^{16}$O and $^{18}$O samples. Furthermore, we compared the thermal Hall angle ratio of our samples by plotting $|\kappa_{xy}/\kappa_{xx}|$ as a function of temperature. Interestingly, while the thermal Hall angle ratio of the STO16 sample increases upon cooling, the thermal Hall angle ratio of the STO18 samples shows a significant drop below 25 K (see Fig. 4b). This difference in the thermal Hall angle ratio for the two types of samples again indicates that the mechanism is different at STO16 and STO18 below 25 K, which we note to be the ferroelectric transition temperature. It is also worthwhile noting that there are no visible thermal cycling effects in our THE measured STO18 when warmed above the AFD transition temperature, as shown in Fig. 4c.

We would now like to discuss the origin of the isotope effects observed in the thermal Hall conductivity of SrTiO$_3$. The two recent papers, first experiment [12], and second theory [27], claimed that the large THE found in their SrTi$^{16}$O$_3$ measurements is most likely to be due to structural domain effects, and associated with the large dielectric constant. Especially, Ref. [12] suggested that the two orders of magnitude smaller $\kappa_{xy}$ in KTaO$_3$ comes from the absence of AFD structural transition even if KTaO$_3$ is also a quantum paraelectric comparable to SrTi$^{16}$O$_3$. However, the recent report of a THE in cuprate Mott insulators questioned the contribution of AFD structural domains in SrTi$^{16}$O$_3$ by showing that other tetragonal perovskites such as Nd$_2$CuO$_4$ and Sr$_2$CuO$_2$Cl$_2$ without AFD structural domains have large $\kappa_{xy}$ values in comparison to SrTi$^{16}$O$_3$ [30]. Moreover, we found the different behavior between STO16 and STO18 samples, and STO16 and KTaO$_3$, although our STO18 undergoes the same AFD transition - as seen by the plateau in the longitudinal thermal conductivity above 100 K - as STO16, and KTaO$_3$ does not [28, 34]. Therefore, there should not be any difference between our two samples from the viewpoint of the AFD structural domain scenario put forward in Ref. [12]. That there is no visible thermal hysteresis in our STO18 is further evidence against the AFD structural domain scenario.

Instead, we should point out that there is a significant drop in the Hall angle for STO18 below the ferroelectric transition temperature. It is to be noted too that the peak positions appear at different temperatures for $\kappa_{xx}$ and $\kappa_{xy}$ for STO18, unlike STO16. Such behavior of thermal Hall conductivity is consistent with quantum fluctuations suppressed in STO18 by oxygen exchange, eventually leading to a ferroelectric state as in Ref. [24]. Thus, in the ferroelectric state of STO18, there cannot be any quantum fluctuations remaining. This observation appears to be consistent with what it is found in $KTaO_3$ in Ref. [12]. $KTaO_3$ is out of the low-temperature quantum critical paraelectric region and is much further from the quantum critical point on the phase diagram than STO16, as one can see in Ref. [19]. Therefore, quantum fluctuations are far more dominant in STO16. Due to the simplicity of the system, and the few differences between STO18 and STO16, we can certainly conclude that whatever the underlying mechanism is, it should be related to the transverse optical phonon mode strongly affected by the oxygen masses. We feel it is perhaps more fitting to regard STO16 as having a greatly enhanced thermal Hall signal, rather than STO18 and $KTaO_3$ having diminished ones, and that this enhancement is linked to the presence of quantum critical fluctuations of the polarization and phonons, rather than structural domains.

In summary, we have carried out thermal Hall conductivity measurements on two types of samples of $SrTi^{16}O_3$ and $SrTi^{18}O_3$. Our experimental results demonstrate a significant isotope effect, although both show almost equivalent, qualitatively and quantitatively, temperature dependence of the longitudinal thermal conductivity. The isotope effect entails the drastic reduction in the thermal Hall effect and the completely different temperature dependencies in the thermal Hall angle for $SrTi^{16}O_3$ and $SrTi^{18}O_3$. Our experimental observations demonstrate beyond a reasonable doubt that whatever the microscopic mechanism may be that lies behind the THE of $SrTiO_3$, it ought to be of intrinsic origin and cannot be due to the presence of antiferrodistortive structural domains. It is most likely linked to the transverse optical phonon mode increasingly evidenced as fundamental to both quantum critical phases in $SrTi^{16}O_3$ and the enigmatic dilute superconductivity in electron-doped $SrTi^{16}O_3$.

**Acknowledgments**
We would like to thank Jaehong Jeong for helpful comments. Work at CQM was supported by the Leading Researcher Program of the National Research Foundation of Korea (Grant No. 2020R1A3B2079375). Work at IBS CCES was supported by the Institute for Basic Science of the Republic of Korea (Grant No. IBS-R009-G1). M.I. was supported by MEXT Elements Strategy Initiative to Form a Core Research Center of Japan.

**References**
[1] N. W. Ashcroft & N. D. Mermin, *Solid State Physics* (Holt, Rinehart and Winston, New York, 1976).
[2] H. Katsura, et al., *Phys. Rev. Lett.* **104**, 066403 (2010).
[3] C. Strohm, et al., *Phys. Rev. Lett.* **95**, 155901 (2005).
[4] Y. Onose, et al., *Science* **329**, 297 (2010).
[5] M. Hirschberger, et al., *Science* **348**, 106 (2015).
[6] M. Hirschberger, et al., *Phys. Rev. Lett.* **115**, 106603 (2015).
[7] T. Ideue, et al., *Nature Materials* **16**, 797 (2017).
[8] K. Sugii, et al., *Phys. Rev. Lett.* **118**, 145902 (2017).
[9] Y. Kasahara, et al., *Nature* **559**, 227 (2018).
[10] Y. Kasahara, et al., *Phys. Rev. Lett.* **120**, 217205 (2018).
[11] G. Grissonnanche, et al., *Nature* **571**, 376 (2019).
[12] X. Li, et al., *Phys. Rev. Lett.* **124**, 105901 (2020).
[13] H. Ishizuka and N. Nagaosa, *Sci. Adv.* **4**, eaap9962 (2018).
[14] R. Chisnell, et al., *Phys. Rev. Lett.* **115**, 147201 (2015).
[15] J. H. Han and H. Lee, *J. Phys. Soc. Jpn.* **86**, 011007 (2017).
[16] K. Hwang, et al., arXiv:1712.08170v1.
[17] H. Lee, et al., *Phys. Rev. B* **91**, 125413 (2015).
[18] G. Grissonnanche, et al., arXiv:2003.00111.
[19] S. E. Rowley, et al., *Nature Physics* **10**, 367–372 (2014).
[20] M. J. Coak, et al., *PNAS* **117**, 12707-12712 (2020).
[21] Y. Tomioka, et al., *Nature Communications* **10**, 738 (2019).
[22] J. Ruhman and P. A. Lee, *Phys. Rev. B* **94**, 224515 (2016).
[23] M. Itoh, et al., *Phys. Rev. Lett.* **82**, 3540 (1999).
[24] R. Wang and M. Itoh, *Phys. Rev. B* **64**, 174104 (2001).
[25] H-L. Kim, et al., *Rev. Sci. Instr.* **90**, 103904 (2019).
[26] Y. Yamada and G. Shirane, *J. Phys. Soc. Jpn.* **26**, 396-403 (1969).
[27] J. Chen, et al., *Phys. Rev. Lett.* **124**, 167601 (2020).
[28] V. Martelli, et al., *Phys. Rev. Lett.* **120**, 125901 (2018).
[29] B. K. Agarwalla, et al., *Eur. Phys. J. B* **81**, 197-202 (2011).
[30] M. Boulanger, et al., arXiv:2007.05088.
[31] L. Sheng, et al., *Phys. Rev. Lett.* **96** 155901 (2006).
[32] M. Mori, et al., *Phys. Rev. Lett.* **113** 265901 (2014).
[33] T. Saito, et al., *Phys. Rev. Lett.* **123** 255901 (2019).
[34] E. F. Steigmeier, *Phys. Rev.* **168** 523 (1968).

**Figures**

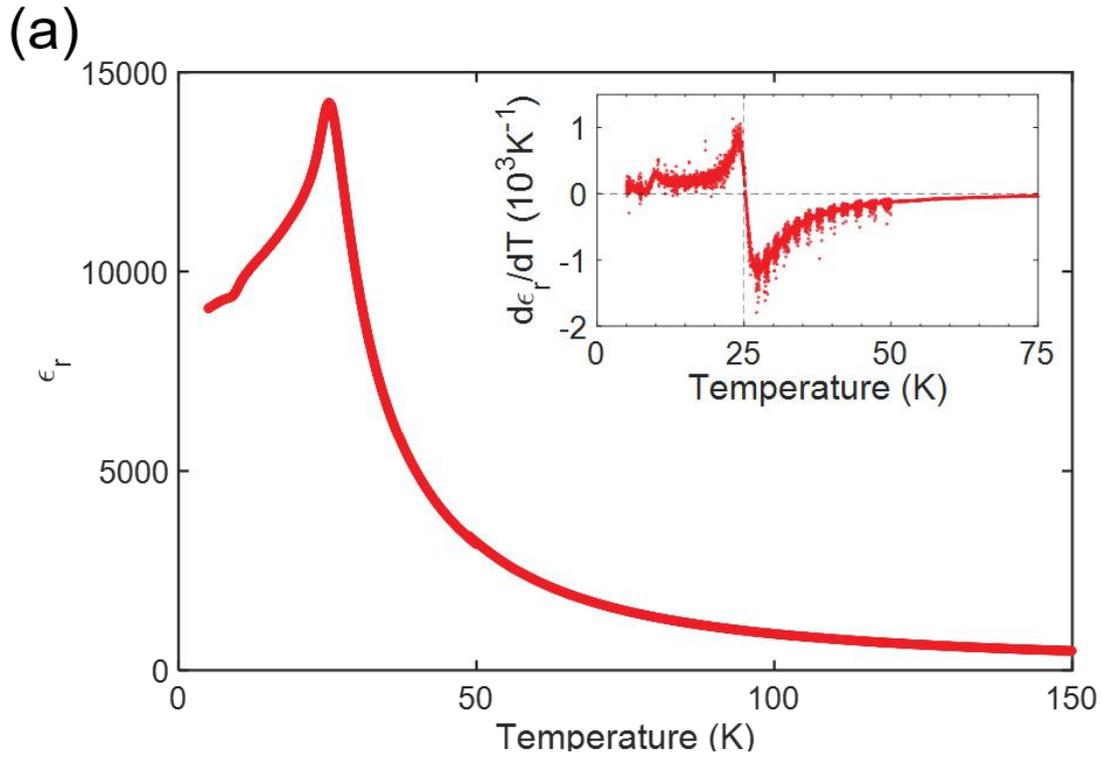

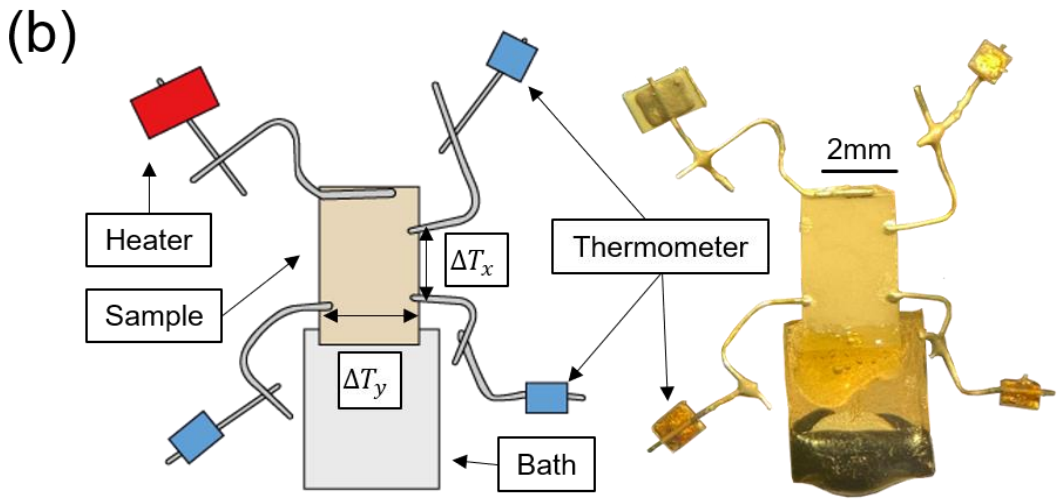

**Fig. 1** (a) Dielectric constant of SrTi$^{18}$O$_3$ sample used for thermal hall measurement. Its peak position is located near 25 K, which is consistent with a previous study in Ref. [19]. (b) Schematic figure and photo of the thermal Hall measurement setup. It consists of one heater and three thermometers with a sample loaded on a sapphire cold finger.

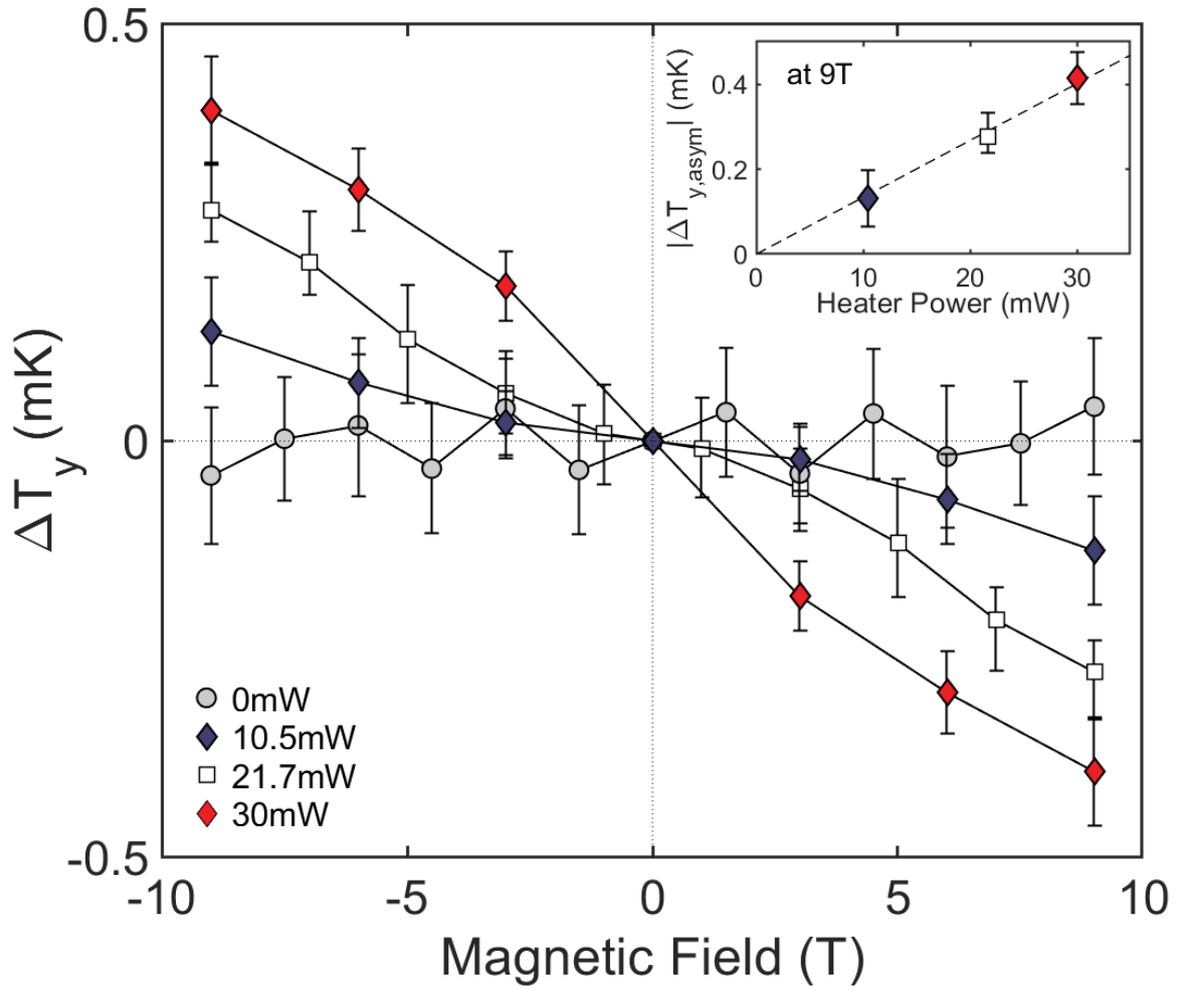

**Fig. 2** Anti-symmetrized transverse temperature difference with different heater power applied (thermal gradient), measured on one of our SrTi$^{18}$O$_3$ (STO18 #2) samples. It exhibits good linearity to heater power (see the inset) and applied field, and a distinct difference from the noise level (0 mW). All measured at an average sample temperature of 40 K.

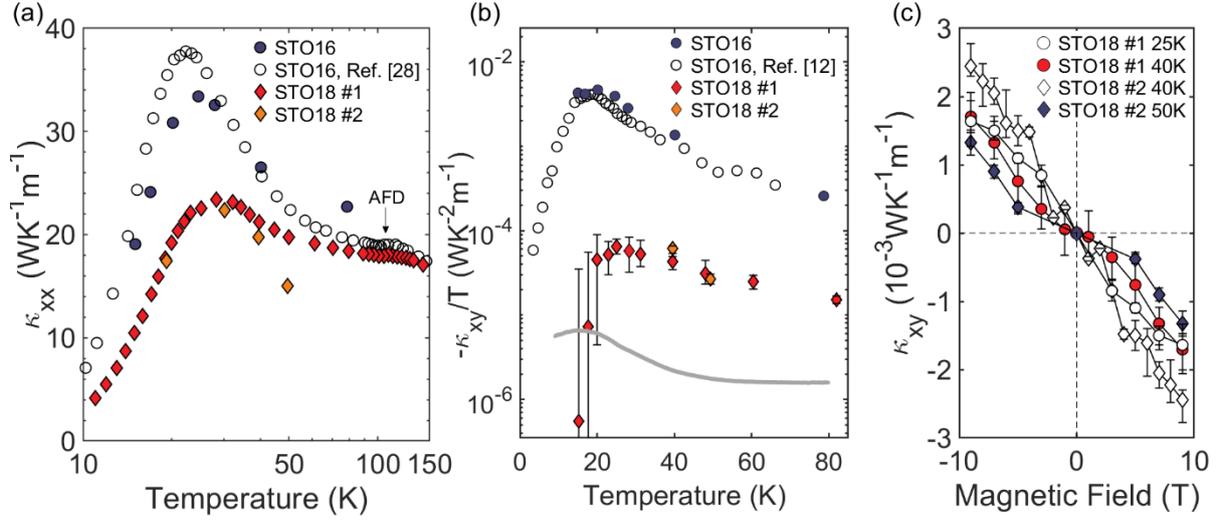

**Fig. 3** (a) Longitudinal thermal conductivity $\kappa_{xx}$ of SrTi$^{16}$O$_3$ (STO16) and SrTi$^{18}$O$_3$ (STO18). STO18 samples have a slightly lower $\kappa_{xx}$ value but behave similarly to STO16. Both have similar peak positions and a subtle suppression of $\kappa_{xx}$ near the AFD transition around 110 K. Error bars are smaller than the symbol size. (b) Temperature dependence of thermal Hall conductivity for several SrTiO$_3$ samples. For comparison, we also plot the data from Ref. [12]. The solid grey line indicates the theoretical boundary of the measurable thermal Hall conductivity using our setup [25]. (c) Field dependence of thermal Hall conductivity. The proportionality is consistent over the measured temperature range.

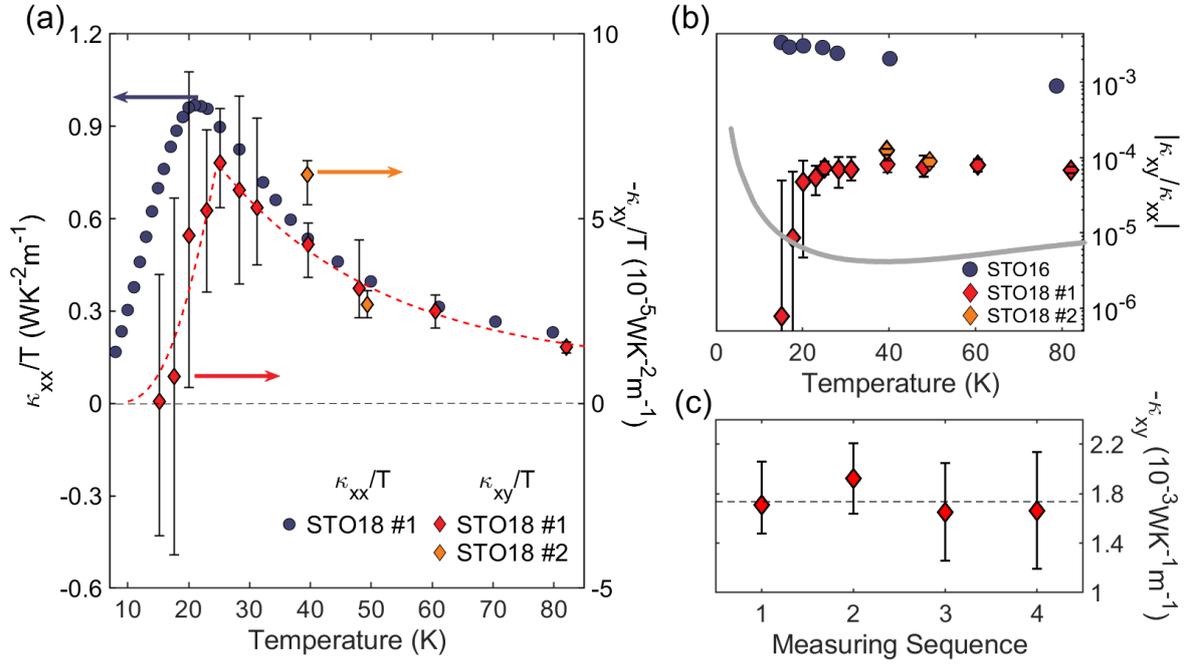

**Fig. 4** (a) Thermal Hall conductivity SrTi$^{18}$O$_3$ (STO18) divided by temperature, plotted against temperature. Unlike in SrTi$^{16}$O$_3$ (STO16), their peak locations do not coincide. (b) The thermal Hall angle ratio plotted against temperature. Continuing from (a), the thermal Hall angles differ significantly in their order, along with a rapid decrease below 25 K. The solid grey line indicates the theoretical boundary of the measurable thermal Hall angle ratio using our setup [25]. (c) Repeated measurement of thermal Hall conductivity at 40 K with magnetic field 9 T. The sample was warmed up to room temperature (above the AFD transition temperature) between each sequence.